\documentclass[fleqn,10pt]{wlscirep}
\usepackage[utf8]{inputenc}
\usepackage[T1]{fontenc}
\usepackage{bm}

\usepackage{pdfpages}

\usepackage{graphicx}
\usepackage{tcolorbox}
\usepackage{xcolor}

\usepackage{stfloats}
\usepackage{float}
\usepackage{tcolorbox}
\usepackage{xcolor}%
\usepackage{placeins}

\title{Climate Change in the European Political Debate: Testing the Incongruence between Elite Agendas and Audience Engagement}

\author[1,+]{Irene Scalco}
\author[2,*]{Matteo Cinelli}

\affil[1]{Department of Social Sciences and Economics, Sapienza University of Rome, P.le Aldo Moro, 5, Rome, 00185, Italy}

\affil[2]{Department of Computer Science, Sapienza University of Rome, Viale Regina Elena 295, Rome, 00161, Italy}

\affil[+]{irene.scalco@uniroma1.it}
\affil[*]{matteo.cinelli@uniroma1.it -- Corresponding author}

\begin{abstract}
Climate change ranks among European citizens' major concerns, but how strongly is this concern reflected in political debate? We examine online political communication surrounding the 2024 European Parliament elections across seven countries, analysing 27,650 Facebook posts from 64 political parties. We perform multilingual classification combining an agentic AI pipeline with network-based topic modelling to investigate two dimensions of climate communication. First, we compare the prevalence of posts on climate and environmental issues with that of posts on the economy, migration, civil rights, and security and defence. Climate and environmental issues appear in 11.7\% of classifiable posts, making them the least frequently addressed of the five categories. Coverage is substantially higher among left and centre-left parties than among centre-right and right-wing parties. Second, we test for incongruence between elite agendas and audience engagement by comparing the parties' thematic coverage with the distribution of engagement across topical clusters of posts. The resulting differences vary across parties and themes, thus showing the presence of an incongruence that, however, shows no clear ideological gradient.
\end{abstract}
\begin{document}

\flushbottom
\maketitle

\thispagestyle{empty}

\section {Introduction}

%Context: 
``What do you think are the three biggest challenges that people in your country are currently facing?''. This is one of the questions included in the Climate Survey conducted by the European Investment Bank (EIB) \cite{EIBClimateSurvey}. In the 2023–2024 survey, covering more than 30,000 respondents, participants could select up to three challenges from a range of issues, including the rising cost of living, income inequalities, unemployment, climate change, terrorism, cyberattacks, large-scale migration, political instability, access to healthcare, and environmental degradation. Climate change was the second most frequently selected challenge in Europe, placing it among the issues that citizens perceived as particularly important. The survey also coincided with an exceptional electoral period. The year 2024 was widely described as a ``year of global elections'' \cite{TIMEElectionYear}, with elections taking place in countries representing a large share of the world's population \cite{UNDP2024}, including the European Parliament elections. This electoral context raises an important question: to what extent was public concern about climate change reflected in political parties' communication during the electoral campaign?

Public concern does not necessarily translate into political attention \cite{hartland2026role, kluver2016responds}. During electoral campaigns, political actors decide which issues to prioritize \cite{gilardi2022social, kluver2016setting} and how to frame them, while different topics compete for limited public attention \cite{lorenz2019accelerating, weng2012competition}. Political parties, as political-elite actors, can provide cues that associate policy issues with particular political positions and identities \cite{van2021elite}. The topics they select and the attention they dedicate to them are therefore important dimensions of political communication. Social media platforms have become an important arena in which these dynamics can be observed \cite{schafer2026social, loru2025ideology}. They provide political parties with direct channels to communicate their priorities, while allowing users to selectively engage with the content they receive \cite{cinelli2021echo, cinelli2020selective}. However, the issues emphasized by political actors are not necessarily those that receive the greatest attention from their audiences \cite{ennser2022does}. This distinction is also relevant to research on political polarization, which shows that elite and mass dynamics can develop and contribute differently to political divisions \cite{falkenberg2026toward, salloum2025anatomy, rinscheid2021shapes}. Political communication and audience response are therefore connected but distinct dimensions of online debate. Examining both can help distinguish the priorities expressed by political actors from the attention their communication receives.

Climate change provides a particularly useful context for examining these dynamics. It is one of the major challenges of this century \cite{IPCC2022} and a multidimensional issue \cite{misuraca2026energy} that intersects with economic policy, energy, environmental protection, social justice, migration, agriculture, and technological innovation \cite{scalco2026modelling, chen2021polarization}.
In this context, the extent to which climate change is discussed and which aspects of the issue receive attention become particularly relevant.
Political actors may indeed differ both in the overall attention they dedicate to climate change \cite{farstad2018explains} and in the specific aspects they emphasize. Previous research on online climate communication shows that attention to climate change is not constant, but responds to specific events such as extreme weather events \cite{meyer2023between, torricelli2023does, roxburgh2019characterising}, international climate conferences, and climate protests \cite{lochner2024climate}. At the same time, the online climate debate can be characterized by ideological polarization \cite{falkenberg2022growing, chen2021polarization} and by the circulation of misleading or unreliable information \cite{treen2020online, van2017inoculating, storani2025relative}.
Political parties may prioritize certain aspects in their communication \cite{hart2025climate, cohen2003party}, while their audiences may engage more strongly with others \cite{sangiorgio2025evaluating}. A party may therefore devote considerable attention to climate change while receiving a disproportionate share of engagement on different themes. Conversely, limited climate coverage may coexist with a similar distribution of thematic emphasis and engagement.

This distinction raises a question that extends beyond whether left-leaning parties discuss climate change more frequently than right-leaning parties: do ideological differences in climate salience also correspond to systematic differences in the coverage, related themes and audience engagement

%Contributo: 
We address this question by examining political parties' Facebook communication across seven European countries -- Austria, Denmark, France, Germany, Italy, Spain, and Sweden -- ahead of the 2024 European Parliament elections. We first assess the coverage of climate and environmental issues relative to four other broad categories: economy, migration, civil rights, and security and defence. We then reconstruct the thematic structure of climate-related posts and compare each party's distribution of thematic coverage with the corresponding distribution of engagement. This approach connects the allocation of political attention across issues with users' engagement across themes within climate-related communication. We treat political parties' Facebook posts as a proxy for the supply of political content \cite{scalco2026detect} and engagement with these posts as an indicator of audience attention. 
%Engagement captures an observable response to the content supplied by political actors, rather than directly measuring users' preferences, agreement, or political opinions. 
Our analysis examines where political coverage and audience attention converge or diverge across climate-related themes, interpreting differences as elite-mass dynamics that can be interpreted as signals of category incongruence \cite{yoda2026time, shim2020measuring}.

%Our analysis covers seven European countries: Austria, Denmark, France, Germany, Italy, Spain, and Sweden. 
We analyse Facebook posts published by 64 political parties and 133 news outlets, covering 27,650 party posts and 1,289,398 news posts between September 2023 and May 2024. Rather than considering climate change in isolation, we compare its coverage with four other broad areas of political debate: \textit{economy}, \textit{migration}, \textit{civil rights}, and \textit{security and defence}. Specifically, we ask: (i) how prominent are climate and environmental issues relative to other major issues in the communication of political parties on Facebook? (ii) how do their prominence and thematic composition vary between political orientations and national contexts? and (iii) do ideological differences in climate importance correspond to a systematic pattern in the divergence between thematic coverage of parties and audience engagement?

%Metodo: 
To address these questions, we first develop an agentic topic-modelling pipeline \cite{lan2024stance} based on five independent zero-shot LLM agents, each specialized in one of the five broad topics, to determine whether a post addresses a given issue. The classification framework is described in detail in Sections \ref{sec:Topic_selection} and \ref{sec:topic_modelling_pipeline}. This first step allows us to compare the presence of climate change and other political issues across parties, political orientations, and countries. We then focus specifically on climate-related communication. Using a second LLM-based procedure, we extract the main keywords from climate-related posts and construct a party--keyword network. We apply a hierarchical Stochastic Block Model (hSBM) \cite{gerlach2018network} to this network to identify groups of related keywords and reconstruct the thematic structure of the climate debate. This data-driven approach allows the main climate topics to emerge from the content itself, rather than defining them in advance. We then use this thematic structure to compare the topics emphasized by political parties with the engagement they receive from their audiences.

%Findings: 
Three main findings emerge from our analysis. First, despite being among citizens' main stated concerns, climate change occupies a relatively limited position in online political communication. Among the five broad topics considered, it is the least frequently addressed by political parties and also receives comparatively limited coverage from news outlets. Second, attention to climate change varies across political orientations. Left and centre-left parties dedicate greater coverage to climate-related issues, while their coverage is substantially lower among centre-right and right-wing parties. Similar patterns emerge from the broader topic profiles of political parties, with ideologically related parties showing comparable agendas even across different national contexts. Third, within the climate debate, the relationship between political coverage and audience attention is nuanced. Differences vary across topics and parties and are generally limited, with statistically significant gaps concentrated in a relatively small number of cases. Rather than pointing to a systematic disconnect between political actors and their audiences, these results suggest that alignment depends on both who is communicating and which aspect of climate change is being discussed.

%Organizzazione: 
The paper is organized as follows. Section \ref{sec:data_and_methods} presents the dataset and analytical framework. Section \ref{sec:results_and_discussion} presents the results, moving from the position of climate change within the broader political agenda to its internal thematic structure and the relationship between political emphasis and audience attention. Section \ref{sec:conclusion} discusses the implications of these findings for climate communication and political polarization.

\section{Data and Methods} \label{sec:data_and_methods}
\subsection{Dataset}\label{sec:dataset}
We focus on seven European countries participating in the 2024 European Parliament elections: Austria, Denmark, France, Germany, Italy, Spain, and Sweden. The selection is based on the European Sustainable Development Report 2026, which ranks countries according to their performance across the 17 Sustainable Development Goals (SDGs) \cite{lafortune2026europe}. The sample includes three top-ranked countries (Denmark, Sweden, and Austria), two middle-ranked countries (Germany, France), and two lower-ranked countries (Italy and Spain). For each country, we manually compile a list of Facebook pages associated with the official accounts of political parties represented in the national parliaments, as well as major national news agencies. The political leaning of each party is manually labelled using its corresponding Wikipedia page as a reference.

The full dataset includes 64 political parties and 133 news outlets across the seven countries. It contains 27,650 posts from political parties and 1,289,398 posts from news outlets collected between September 1, 2023, and May 1, 2024. For each post, we record the page name, posting time, textual content, and aggregated engagement metrics, including the number of reactions and comments. Data are collected through CrowdTangle (now Meta Content Library API), a public insights tool operated by Meta Platforms, and the analysis is conducted in compliance with its terms and conditions. A complete list of the pages included in the analysis is provided in the SI. CrowdTangle was a Meta-owned data collection tool that remained active until August 14, 2024. The tool provided access to a limited set of public accounts. For Facebook, the available dataset included more than 7 million public Facebook pages and verified profiles. Specifically, it covered all public Facebook pages with more than 50,000 likes, all public groups with more than 95,000 members, all US-based public groups with more than 2,000 members, and all verified profiles.

We focus on Facebook because of its widespread use across the seven European countries during the period under study. According to the ``Digital Services Act Transparency Report'', Facebook had an average of 4.2 million monthly active users in Austria, 4.4 million in Denmark, 42.4 million in France, 33.3 million in Germany, 35.9 million in Italy, 26.9 million in Spain, and 7.2 million in Sweden \cite{Meta_Report}. 

\subsection{Topic selection} \label{sec:Topic_selection}
The first step in the analysis is to identify a set of broad topics that can be used to assess the relative prominence of climate-related discourse within the wider political debate during the 2024 European Parliament election period. Rather than examining climate change alone, we compare it with other major issues that were relevant to the European and national political agendas during the period under study. This approach allows us to assess whether climate-related issues were prominent relative to other topics competing for public and political attention. To define the thematic categories, we use the ``Topics'' section of the European Parliament website \cite{europarl_topics} as a starting point and complement it with issues that were particularly salient in the political debate during the period under analysis. We identify five broad topics: \textit{Climate Change}, including climate and environmental issues, biodiversity, and the climate transition; \textit{Economy}, including economic recovery, taxation, and budgetary issues; \textit{Migration}, covering migration-related issues; \textit{Civil Rights}, including gender equality, social protection, education, youth, and cultural issues; and \textit{Security}, including terrorism and other security-related issues.
%These five topics cover broad areas of political debate that are relevant both at the European and national levels. Their use therefore provides a common framework for comparing the relative prominence of climate-related discourse across countries while accounting for the different issues competing for attention during the election period.

\subsection{Process of Topic Modelling} \label{sec:topic_modelling_pipeline}
Once the topics are defined, the next step is to classify the posts according to these thematic categories. To do so, we develop a topic classification pipeline based on large language models (LLMs), using Gemma 3 4B, a model from Google’s Gemma family. We select this model for its multilingual capabilities, as it was trained on data covering more than 140 languages \cite{team2025gemma}, allowing us to process posts in the six languages included in our sample: German, Danish, French, Italian, Spanish, and Swedish.
The model takes as input the textual content of all posts. The classification is performed separately for each country, distinguishing between posts published by political parties and those published by news outlets. For each of the five topics, we define a separate expert agent using a zero-shot approach. The system prompt provided to each agent is: \textit{“You are an expert in $\langle$topic\_name$\rangle$ communication and $\langle$country\_name$\rangle$ political communication on social media”}. The temperature parameter is set to 0 to ensure consistent outputs. The prompt is the same in structure across all five agents and consists of three components. The complete prompt used for the \textit{Climate Change} classification is reported in Figure~\ref{fig:zeroshot-prompt} while prompts for all other topics are available in the SI. The first provides general instructions about the role of the model and the classification task. The second defines the decision rules used to determine whether a post belongs to the topic. The third specifies the required output format.
\begin{figure}
\begin{tcolorbox}[
  colback=gray!5,
  colframe=black!60,
  boxrule=1pt,
  arc=4mm,
  left=8pt, right=8pt, top=8pt, bottom=8pt,
  width=0.95\linewidth]

\small
You are an expert in climate-change communication and $\langle$country\_name$\rangle$ political communication on social media.
\vspace{4pt}

\textbf{Task:}\\
Classify the content of the provided $\langle$country\_name$\rangle$ social-media post.
\vspace{4pt}

\textbf{Decision Rule:}\\
Output 1 ONLY if the post explicitly and substantively addresses real climate-change or environmental issues. These include (but are not limited to) global warming, pollution, renewable energy, sustainability, biodiversity, environmental protection, deforestation, CO\textsubscript{2} emissions, recycling, extreme weather events, climate-related natural disasters, or public policies addressing environmental problems.
Output 0 if the post DOES NOT explicitly discuss any of these topics.
\vspace{4pt}
\textbf{Exclude} (always output 0 when these occur):
\begin{itemize}
    \item Mentions of politicians, parties, political events, or social issues unrelated to environment or climate.
    \item Metaphorical, symbolic, idiomatic, emotional, or figurative uses of words such as climate, green, environment, air, hot, cold, etc.
    \item Any phrase that uses environmental terminology without referring to actual environmental or climate phenomena (e.g., ``political climate'', ``a new air'', ``going full energy'').
    \item Any expression that could be interpreted metaphorically unless the environmental meaning is explicit.
\end{itemize}
\end{tcolorbox}
\caption{Prompt used to classify social-media posts on climate-change content.}
\label{fig:zeroshot-prompt}
\end{figure}
%\FloatBarrier
%

\begin{figure}[H]
    \centering
    \includegraphics[width=1\textwidth]{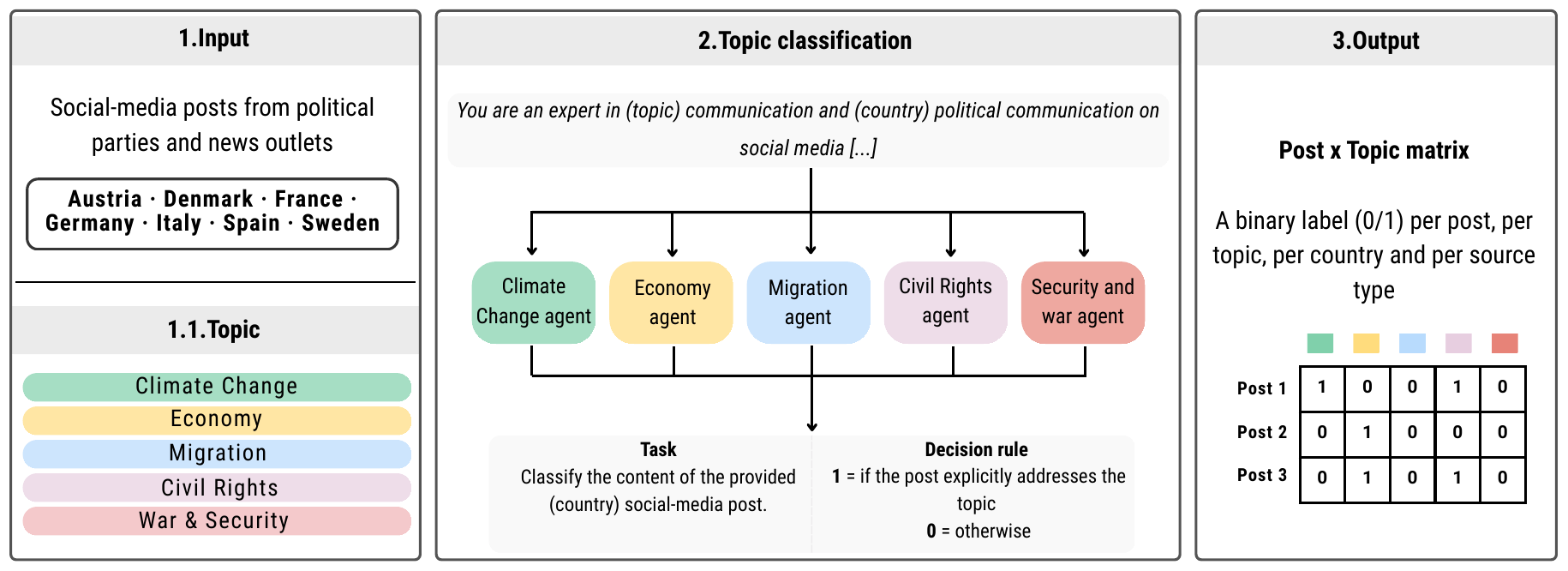}
    \caption{LLM-based topic-modelling pipeline. The pipeline consists of two main steps: (1) Input, in which the countries and topics of interest are defined; and (2) Topic classification, in which a specific prompt is developed and applied to each country–topic combination. The resulting output (3) is a post-by-topic matrix containing binary labels: 1 if a post explicitly addresses a given topic and 0 otherwise.} 
    \label{fig:FigPip}
\end{figure}

Figure~\ref{fig:FigPip} shows the complete LLM-based topic-modelling pipeline. Following a binary classification approach \cite{lan2024stance}, each agent assigns a value of 1 if a post addresses its corresponding topic and 0 otherwise. This approach allows posts to be associated with multiple topics while also permitting an all-zero classification, since the five selected topics do not necessarily cover the entire political debate. Posts receiving a value of 0 from all agents are therefore classified as addressing none of the predefined topics. Further details on the distribution of posts across topics are provided in Section ~\ref{sec:dataset}.

\subsection{Keywords Extraction and Climate Topic Definition} \label{sec:keywords_ext}
In the second part of the project, we focus exclusively on the \textit{Climate Change} topic. We first select all posts classified with a value of 1 in the climate topic column. We then apply a keyword extraction pipeline to this subset, processing the posts separately for each country. Using the same model adopted for topic classification (Gemma 3 4B), we define a dedicated agent with expertise in climate change. The agent reads each post and extracts three to four keywords that capture its main content. The prompts used for each country are reported in the Supplementary Information (SI). Since posts can be assigned to multiple topics, the climate-related subset includes both posts exclusively addressing climate change and posts discussing climate change alongside other issues.
We use the extracted set of keywords to construct a bipartite network linking individual political party pages to the keywords identified in their posts. Before constructing the network, we apply two filtering steps. First, we remove keywords that appear directly in the extraction prompt, namely \textit{climate change}, \textit{social media}, and \textit{political discourse}, to avoid trivial or prompt-induced associations. Second, we remove keywords that occur only once, reducing noise and improving the robustness of the resulting network structure.

We then apply a hierarchical stochastic block model (hSBM) to the filtered bipartite network in order to extract group of keywords that can be interpreted as topics \cite{gerlach2018network}. We select level 0 of the hierarchy, which provides the highest level of granularity. This level identifies 39 blocks in total, of which 21 correspond to distinct keyword-based topic clusters. These clusters are derived from 889 unique keywords and provide a more detailed view of the main themes that structure political debate around climate change.

\subsection{Measuring the Distance Between Elite Focus and Mass Attention} \label{sec:elite_mass_metrica}
Following the literature on elite--mass dynamics in political communication, we distinguish between political elites, including political parties, elected representatives, and policymakers, and the mass public, broadly understood as ordinary citizens \cite{van2021elite}. In the context of social media, this distinction allows us to examine political communication from the perspective of both the actors producing content and the audiences engaging with it \cite{barbera2019leads}. In our analysis, we operationalize elites as political parties publishing content on Facebook, while the mass dimension is represented by users' engagement with these posts. Accordingly, \emph{elite focus} refers to the thematic coverage expressed in political parties' posts, whereas \emph{mass attention} captures the observable audience response to this content, rather than a representative measure of citizens' preferences.

\paragraph{Topic assignment and post-level topic shares.}
Each keyword extracted from a post is assigned to one of $T$ macro-topics via a hierarchical stochastic block model (hSBM). Let $n_{p,t}$ denote the number of keywords in post $p$ assigned to topic $t$, and let $N_p = \sum_{t=1}^{T} n_{p,t}$ denote the total number of extracted keywords in post $p$. The \emph{topic share} of post $p$ for topic $t$ is defined as the proportion of that post's keywords belonging to topic $t$:
\begin{equation}
\mathrm{TS}_{p,t} = \frac{n_{p,t}}{N_p}, \qquad \sum_{t=1}^{T} \mathrm{TS}_{p,t} = 1.
\end{equation}

\paragraph{Elite focus.}
Let $\mathcal{P}_j$ denote the set of posts published by party $j$, with $|\mathcal{P}_j|$ its cardinality. \emph{Elite focus} ($\mathrm{EF}$) measures the average topic share a party places across its posts, computed as the mean of post-level topic shares:
\begin{equation}
\widetilde{\mathrm{EF}}_{j,t} = \frac{1}{|\mathcal{P}_j|} \sum_{p \in \mathcal{P}_j} \mathrm{TS}_{p,t}.
\end{equation}
To ensure comparability across parties and topics, we normalize within each party so that elite focus shares sum to unity:
\begin{equation}
\mathrm{EF}_{j,t} = \frac{\widetilde{\mathrm{EF}}_{j,t}}{\displaystyle\sum_{t'=1}^{T} \widetilde{\mathrm{EF}}_{j,t'}}.
\end{equation}

\paragraph{Mass attention.}
\emph{Mass attention} ($\mathrm{MA}$) captures audience responsiveness to party posts. For each post $p$, let $I_p$ denote its total engagement, defined as the sum of all user interactions (comments and reactions, including likes and other reaction types). Mass attention weights the topic share of each post by its share of total party engagement, yielding the engagement-weighted mean topic share:
\begin{equation}
\widetilde{\mathrm{MA}}_{j,t} = \frac{\displaystyle\sum_{p \in \mathcal{P}_j} \mathrm{TS}_{p,t} \cdot I_p}{\displaystyle\sum_{p \in \mathcal{P}_j} I_p}.
\end{equation}
This estimator corresponds to the topic share of a randomly selected interaction directed at party $j$, rather than of a randomly selected post. We again normalize within party:
\begin{equation}
\mathrm{MA}_{j,t} = \frac{\widetilde{\mathrm{MA}}_{j,t}}{\displaystyle\sum_{t'=1}^{T} \widetilde{\mathrm{MA}}_{j,t'}}.
\end{equation}

\paragraph{Elite--mass distance.}
We measure the divergence between elite focus and mass attention for party $j$ on topic $t$ as the log ratio:
\begin{equation}
\mathrm{D}_{j,t} = \log\!\left(\frac{\mathrm{EF}_{j,t}}{\mathrm{MA}_{j,t}}\right)
 \label{eq:log_ratio}
\end{equation}
Because both $\mathrm{EF}_{j,t}$ and $\mathrm{MA}_{j,t}$ are normalized to sum to one within party, $\mathrm{D}_{j,t}$ is well-defined whenever neither quantity is zero (in practice, any topic receiving no keywords from party $j$'s posts yields $\mathrm{EF}_{j,t} = \mathrm{MA}_{j,t} = 0$ and is excluded from the analysis). A positive value of $\mathrm{D}_{j,t}$ indicates that party $j$ allocates more communicative effort to topic $t$ than its audience's engagement would suggest, while a negative value indicates relative audience over-responsiveness to that topic. 

To summarize the overall divergence between elite focus and audience attention, we compute the Total Variation Distance (TVD) for each party $j$:
\begin{equation}
    \operatorname{TVD}_j
    = \frac{1}{2} \sum_{t=1}^{T}
    \left| EF_{j,t} - MA_{j,t} \right|.
    \label{eq:tvd}
\end{equation}
Since both distributions sum to one within each party, TVD ranges from 0 to 1. A value of 0 indicates identical distributions of thematic emphasis and engagement, while larger values indicate greater divergence.

\section{Results and Discussion} \label{sec:results_and_discussion}
\subsection{Characterization of political coverage}
We begin by examining how much coverage climate change receives in political communication compared to other major issues on the political agenda. For comparison, we consider four topics: Economy, Migration, Civil Rights, and Security and Defence. Using the topic-modelling pipeline described in Section~\ref{sec:topic_modelling_pipeline}, we estimate the prevalence of each topic across political parties and countries. Our dataset of political party pages contains 27,650 posts. After applying the topic-modelling pipeline, 16,646 posts are classified as addressing at least one of the topics under analysis, while 8,540 posts concern other themes. The remaining 2,464 posts have missing text (NA) and are therefore excluded from the classification. An advantage of this approach is that it also captures topic co-occurrences, allowing a single post to be assigned to multiple topics. Specifically, 6,816 posts discuss a single topic, 4,383 discuss two topics, 2,759 discuss three topics, 1,765 discuss four topics, and 923 discuss all five topics.

\begin{figure}[H]
    \centering
    \includegraphics[width=0.9\textwidth]{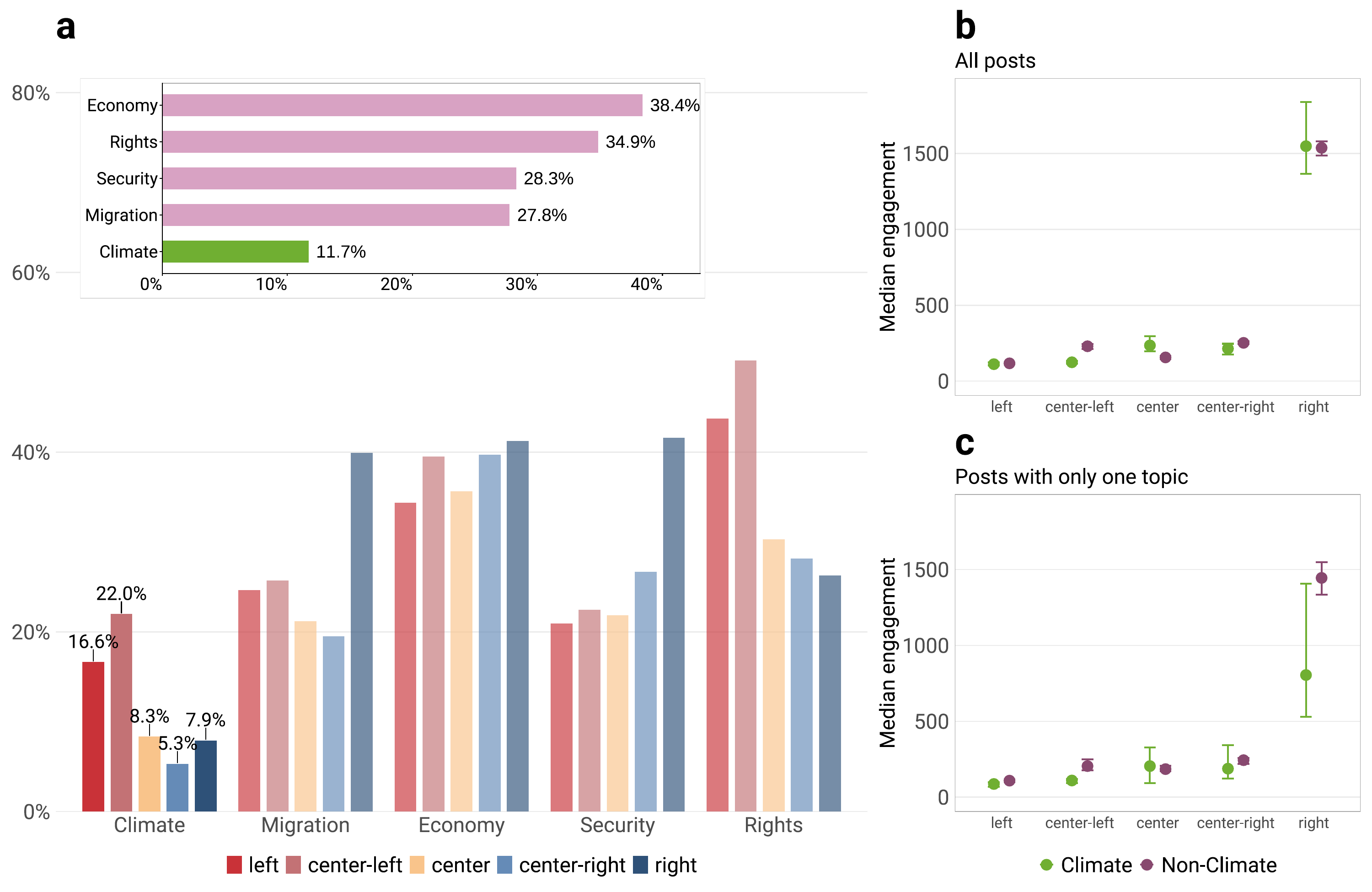}
    \caption{a) The smaller panel shows overall topics coverage. The larger panel shows the distribution of topic coverage across political leanings; the darker bars highlight the climate change topic -- percentages sum to more than 100\% because posts may be assigned to multiple topics. b) Median engagement of all posts, grouped by political leaning. Error bars represent 95\% bootstrap confidence intervals. c) Median engagement of posts assigned to a single topic only, grouped by political leaning. Error bars represent 95\% bootstrap confidence intervals.}
    \label{fig:Fig1}
\end{figure}

As shown in the inset of Figure~\ref{fig:Fig1}(a), climate change receives the lowest level of coverage, in terms of the number of posts published, among all five topics, accounting for 11.7\% of posts regardless of the political orientation of the parties. Economic issues are the topic most frequently discussed across the political spectrum (38.4\%), followed by Civil Rights (34.9\%), Security and Defence (28.3\%), and Migration (27.8\%). These percentages sum to more than 100\% because posts can be assigned to multiple topics. 
Topic rankings vary by political orientation. Civil Rights ranks first among left-wing and center-left parties. Among center-right parties, Economy ranks first, with Civil Rights second and Security and Defence third. Among right-wing parties, Security and Defence ranks first, followed by Economy and Migration.

Despite its prominence in public and policy debates, climate change receives relatively little coverage in electoral communication across the political spectrum. The news outlet data show a similar pattern, with climate change also ranking last among the five topics by post volume. The full results for news outlets are reported in the SI.

Looking more closely at who talks about climate change, figure~\ref{fig:Fig1}(a) shows that center-left and left-wing parties devote the largest shares of their posts to climate change (22\% and 16.6\%, respectively). The corresponding shares are 8.3\% among centrist parties, 7.9\% among right-wing parties, and 5.3\% among center-right parties. This pattern is consistent with expert ratings from the 2024 Chapel Hill Expert Survey (CHES) \cite{rovny20252024}. Many of the parties with the highest climate change salience ratings and the most pro-climate policy positions are left-wing or center-left. In a supplementary analysis of the 54 matched parties, we calculated Spearman correlations between expert-rated climate change salience and the share of climate-related Facebook posts within each ideological group. The strongest correlation was observed among center-left parties ($\rho = 0.82$). The SI presents the party-level scatterplots and the full set of correlations by political orientation.

We also examine which issues parties discuss alongside climate change. In most countries, Economy and Civil Rights are the topics most frequently paired with climate change. The SI reports the full set of pairwise co-occurrence plots by country and political orientation.

To examine whether this pattern extends to user interactions, we compare median engagement between climate-related and non-climate posts within each political orientation. We estimate 95\% bootstrap confidence intervals using 1,000 resamples and conduct two-sided Wilcoxon rank-sum tests, with Holm correction across the ten comparisons. In the multi-topic sample (Figure~\ref{fig:Fig1}b), climate-related posts have lower median engagement in the category center-left and higher median engagement among center parties. Engagement distributions differ significantly in both cases (adjusted $p<0.001$), but no statistically significant differences are detected for the remaining orientations. Among single-topic posts (Figure~\ref{fig:Fig1}c), median engagement is lower for climate-related posts among left, center-left, and right-wing parties. Wilcoxon tests identify significant differences in engagement distributions for these three orientations (adjusted $p=0.025$, $p<0.001$, and $p=0.048$, respectively). Right-wing parties have the largest absolute median gap in this restricted sample. Their climate-only posts receive approximately 44\% lower median engagement than posts assigned exclusively to other topics, whereas the two categories have nearly identical medians in the full sample. Complete test results and additional analyses disaggregated by country and party are provided in the SI.

\subsection{Climate change analysis}
\begin{figure}[H]
    \centering
    \includegraphics[width=1\linewidth]{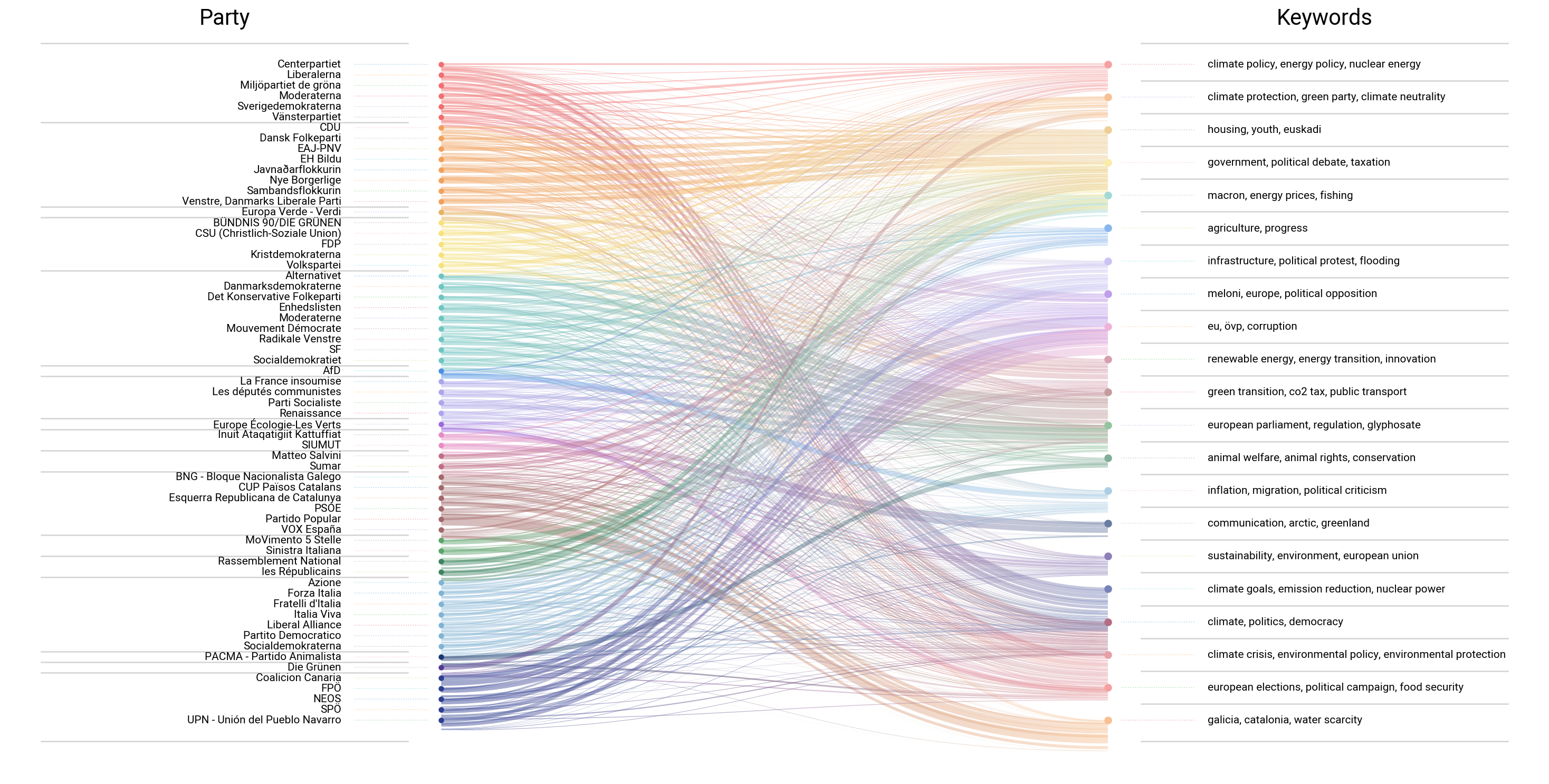}
    \caption{Party–keyword network and thematic clusters identified through the hSBM. Political parties are represented on the left, while keyword clusters are shown on the right, each labelled with three representative keywords. Connections indicate the association between parties and keyword clusters, with colours distinguishing the topics identified by the hSBM.}
    \label{fig:Fig3}
\end{figure}

We conduct an in-depth analysis focused on posts addressing climate change specifically to examine how such a topic is framed across the European political spectrum. We first filter the dataset to retain all posts labelled with a value of 1 for the \textit{Climate Change} topic, including those that also address other issues. This results in a subset of 2,949 posts published by political parties across the seven countries: 762 posts by left-wing parties, 1,007 by center-left parties, 382 by centrist parties, 221 by center-right parties, and 577 by right-wing parties. To characterise the content of these posts, we prompt Gemma 3 4B to extract three to four descriptive keywords in English from each post. These keywords capture the specific issues or actors discussed in the text (see Section~\ref{sec:keywords_ext} for further details). The resulting vocabulary provides a data-driven representation of the main themes present in posts addressing climate change. We then connect the keywords and the parties which produced the posts in a bipartite party/keyword network and apply a hierarchical Stochastic Block Model (hSBM) for topic modelling~\cite{gerlach2018network}. The hSBM identifies groups of related keywords by detecting community structures in the network, without requiring us to define the number of topics in advance. As shown in Figure~\ref{fig:Fig3}, each party (left) is connected to one or more keyword clusters (right), with edge colours indicating topic membership. The results show substantial variation in the themes associated with climate-related political communication. Some parties are connected to a relatively narrow set of themes, such as renewable energy, energy transition, innovation, climate protection, and climate neutrality, while others are associated with a broader range of topics. Notably, keywords such as \textit{EU}, \textit{ÖVP}, and \textit{corruption} also appear alongside environmental themes. This reflects the fact that the analysis includes all posts addressing climate change, including those discussing other political issues. Consequently, the identified clusters capture the broader thematic context in which climate change is discussed, rather than exclusively climate-specific frames. 
\subsubsection{Stochastic Block Model (hSBM) for topic modelling}
We use the clusters identified by the hSBM as sub-topics characterising climate change-related discourse across the European political landscape. We assign a title to each sub-topic based on its constituent keywords, using the same LLM adopted in the previous steps (Gemma 3 4B). We also manually inspect the keyword clusters and their assigned titles to assess their thematic coherence and the appropriateness of the LLM-generated labels. Further details on this process, together with the complete list of keywords and corresponding topic titles, are provided in the SI.

\begin{figure}[H]
    \centering
    \includegraphics[width=1\linewidth]{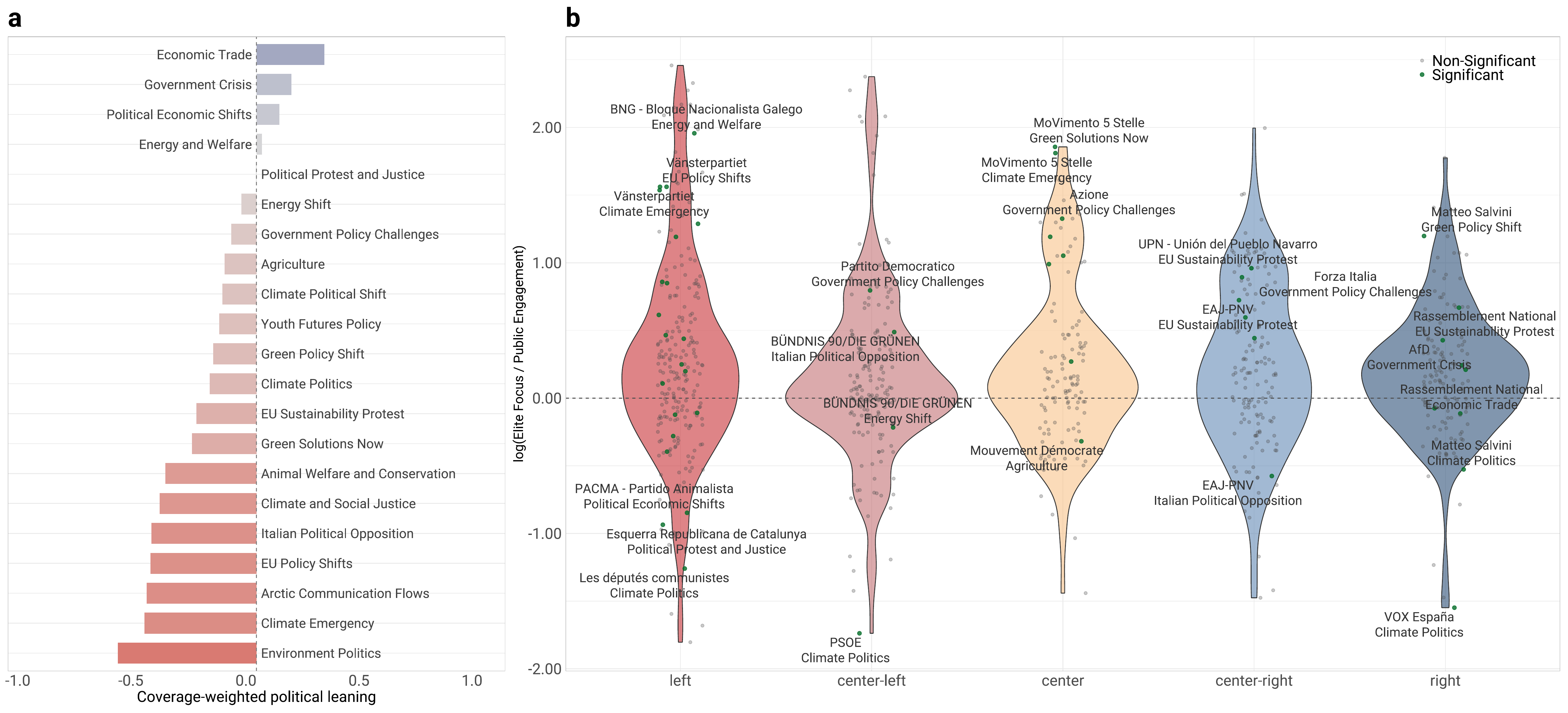}
    \caption{a) Coverage-weighted political leaning of topics identified through the hSBM. Each topic's political leaning is calculated as the average ideological position of the parties associated with it, weighted by the proportion of its keywords used by each party. Negative values indicate stronger associations with left-wing parties, while positive values indicate stronger associations with right-wing parties. b) Distribution of the elite–mass gap across political orientations, measured as the log ratio between elite focus and mass engagement. Positive values indicate that political elites devote relatively more attention to a topic than the engagement it receives, whereas negative values indicate greater mass engagement relative to elite focus. Each point represents a party–topic combination, with green points indicating statistically significant gaps based on permutation tests (p < 0.05).}
    \label{fig:Fig4}
\end{figure}

This classification allows us to examine whether specific topics are more strongly associated with left-leaning or right-leaning parties. Figure~\ref{fig:Fig4}(a) shows the coverage-weighted political leaning of the topics identified through the hSBM. To characterise the political orientation associated with each topic, we first compute the \textit{Topic Keyword Coverage} for each party–topic pair:
\begin{equation}
C_{p,t} = \frac{m_{p,t}}{n_t},
\label{eq:topic_coverage}
\end{equation}
\noindent
where $m_{p,t}$ is the number of different keywords belonging to the topic $t$ used by the party $p$, and $n_t$ is the total number of different keywords assigned to the topic $t$. Then, we assign each political party an ideological score:
\begin{equation}
L_p =
\begin{cases}
-1   & \text{Left}, \\
-0.5 & \text{Center-left}, \\
0    & \text{Center}, \\
0.5  & \text{Center-right}, \\
1    & \text{Right}.
\end{cases}
\label{eq:political_leaning_score}
\end{equation}
For each topic, we calculate the \textit{coverage-weighted political leaning} as:
\begin{equation}
L_t^{(C)} =
\frac{
\displaystyle\sum_{p \in P} C_{p,t} L_p
}{
\displaystyle\sum_{p \in P} C_{p,t}
},
\label{eq:weighted_topic_leaning}
\end{equation}
\noindent
where $P$ denotes the set of parties with an available political leaning score. Parties with zero coverage do not contribute to the weighted average. The resulting score ranges from $-1$ to $+1$. Negative values indicate stronger coverage by left-leaning parties, while positive values indicate stronger coverage by right-leaning parties. Values close to zero indicate little net association with either side of the ideological spectrum.

The results, shown in Figure~\ref{fig:Fig4}, reveal a clear ideological differentiation in the coverage-weighted political leaning of climate-related topics, with most topics showing stronger associations with left-leaning parties. The strongest negative scores are observed for \textit{Environment Politics}, \textit{Climate Emergency}, and \textit{Arctic Communication Flows}, suggesting that their associated vocabulary is more extensively used by left-leaning parties. Similarly, \textit{Climate and Social Justice}, \textit{Animal Welfare and Conservation}, and \textit{EU Policy Shifts} exhibit predominantly left-wing associations. Conversely, only a few topics show positive scores, particularly \textit{Economic Trade}, \textit{Government Crisis}, and \textit{Political Economic Shifts}, indicating stronger associations with right-leaning parties. This suggests that, within posts addressing climate change, right-leaning parties are more strongly associated with vocabulary related to broader economic and political issues. Topics such as \textit{Energy and Welfare} and \textit{Political Protest and Justice} are positioned close to zero, indicating weaker associations with either ideological side.

\subsubsection{Elite focus and Mass attention analysis}

Finally, to study whether the observed differences resonate with the audience of the parties, we examine the distance between elite focus and audience engagement across the identified sub-topics. The metrics used to measure this gap are described in Section~\ref{sec:elite_mass_metrica}.
Figure~\ref{fig:Fig4}(b) summarises the divergence between elite and audience attention across the 21 topics and five political orientations: left, center-left, center, center-right, and right. The panel shows the distribution of party-level log-ratios (Eq. \ref{eq:log_ratio}) within each group, with statistically significant party–topic pairs identified through permutation tests highlighted in dark green. Positive values indicate that parties devote relatively more attention to a topic than the engagement it receives, while negative values indicate that audience engagement exceeds the relative emphasis placed on the topic by political elites.

Across all political orientations, the distributions are largely concentrated around zero, suggesting that elite focus and audience engagement are generally aligned. 

However, substantial variation emerges within each ideological group, with both positive and negative deviations. Left-wing parties display several pronounced positive gaps, particularly for \textit{Energy and Welfare}, \textit{EU Policy Shifts}, and \textit{Animal Welfare and Conservation}, indicating that certain parties devote relatively more attention to these topics than the audience engagement they generate would suggest. Conversely, negative gaps emerge for \textit{Climate Politics} and \textit{Political Protest and Justice}, suggesting comparatively greater audience engagement. Center-left parties show a narrower concentration around zero, although notable negative deviations are observed for \textit{Climate Politics}. Among centrist parties, the largest positive gaps are associated with \textit{Green Solutions Now} and \textit{Climate Emergency}, particularly for Movimento 5 Stelle, while \textit{Agriculture} exhibits a negative gap for Mouvement Démocrate. Center-right parties show both positive and negative deviations, with significant positive gaps for \textit{EU Sustainability Protest} and \textit{Government Policy Challenges}. Finally, right-wing parties display a relatively concentrated distribution around zero, although significant deviations emerge for \textit{Green Policy Shift} and \textit{Climate Politics}, including a pronounced negative gap for VOX España.

Finally, the general alignment emerging from the analysis of Figure \ref{fig:Fig4}b is confirmed also by the values Total Variation Distance (Eq. \ref{eq:tvd}), calculated for each political party across the 21 topics. Indeed, the results show differences across political orientations, without a clear linear ideological gradient. Center-right parties exhibit the highest median distance (0.176), followed by left-wing (0.142), right-wing (0.131), centrist (0.117), and center-left parties (0.080). However, left-wing parties show the highest mean distance (0.187) and the greatest variability (SD = 0.135), with individual values reaching 0.574. In contrast, right-wing parties exhibit the lowest mean distance (0.122) and relatively limited dispersion (SD = 0.054). Overall, these descriptive results suggest that the divergence between political communication and audience engagement varies both across and within ideological groups, without an apparent systematic increase towards the ideological extremes. Additional results on the Total Variation Distance are provided in the SI.

Our findings distinguish two patterns in the ideological organization of climate communication. Political orientation is associated with substantial differences in the share of communication devoted to climate and environmental issues. However, the divergence between thematic emphasis and engagement does not follow an equivalent linear ideological gradient. Greater climate salience therefore does not necessarily coincide with a closer correspondence between the themes a party emphasizes and the distribution of interactions its posts receive.
This distinction helps to clarify ideological differences in climate communication. In this sample, they are evident in the prominence of climate-related issues within party agendas, while differences between thematic emphasis and engagement depend more on individual parties and topics. 
%These results suggest that the factors associated with allocating communication to climate issues may differ from those associated with the distribution of engagement within that communication. Identifying those factors requires further analysis; the present results establish the distinction between the observed patterns.

\section{Conclusions} \label{sec:conclusion}
This study examined the place of climate change in political parties' online communication across seven European countries ahead of the 2024 European Parliament elections. By combining broad issue classification with a detailed analysis of climate-related posts, we investigated two dimensions of political attention: the prominence of climate and environmental issues within party agendas, and the relationship between parties' thematic focus and the engagement their posts receive. Our findings show that climate change received less coverage than the other major political issues considered, despite its importance in public and policy debates. Its prominence also varied across the ideological spectrum, with left-wing and center-left parties devoting substantially more attention to climate-related issues than center-right and right-wing parties. These findings complement research on the variation in the prominence and communication of parties' climate agendas\cite{hart2025climate}. The thematic analysis further shows that climate-related posts encompass economic, social, and political concerns, reinforcing the importance of examining the different dimensions through which environmental issues receive visibility \cite{dubey2025talking, hautea2021showing}. However, these ideological differences in climate coverage do not translate into an equivalent pattern of elite–mass incongruence. We operationalize elite–mass incongruence as the divergence between the thematic focus of political parties and the distribution of engagement their posts receive. Although some party–topic combinations exhibit substantial divergence, the overall distance varies across parties and topics without a clear linear ideological gradient. Greater attention to climate change therefore does not necessarily correspond to a closer alignment between parties' thematic priorities and audience engagement. Moreover, a relatively balanced distribution of thematic coverage and engagement can coexist with limited attention to climate change within the broader political agenda. These findings highlight the importance of distinguishing the prominence of an issue in political communication from the relationship between elite focus and audience response.

Several limitations should be acknowledged. First, the analysis focuses exclusively on Facebook, limiting the generalisability of the findings to other social media platforms. Second, the observation period is restricted to the months preceding the 2024 European Parliament elections, from September 2023 to May 2024. This period provides a relevant context for examining electoral communication, it does not allow us to assess whether the observed patterns persist over longer periods or outside electoral contexts. Third, the analysis covers seven European countries and a selected set of political party pages, meaning that the findings cannot be assumed to represent the entire European political landscape. 

These limitations also indicate directions for future research. Extending the analysis to additional platforms, countries, and longer observation periods would help assess whether the observed patterns are consistent across different political and communication contexts. Furthermore, longitudinal analyses could examine how parties' thematic priorities and audience engagement evolve over time, particularly whether changes in political communication precede or follow shifts in engagement. Overall, this study highlights a distinction in the ideological organisation of climate communication: while political orientation is associated with the prominence of climate change within party agendas, elite–mass incongruence varies more heterogeneously across parties and topics. Examining both dimensions provides a more comprehensive understanding of how climate change enters political agendas and how audiences respond to the content political actors produce.
\bibliography{sample.bib}

@misc{EIBClimateSurvey,
  author = {{European Investment Bank}},
  title = {{EIB Climate Survey: Resources}},
  howpublished = {\url{https://www.eib.org/en/surveys/climate-survey/all-resources}}
}

@book{IPCC2022,
  author = {IPCC},
  title = {Climate Change 2022: Impacts, Adaptation, and Vulnerability},
  editor = {H.-O. Pörtner and D.C. Roberts and M. Tignor and E.S. Poloczanska and K. Mintenbeck and A. Alegría and M. Craig and S. Langsdorf and S. Löschke and V. Möller and A. Okem and B. Rama},
  publisher = {Cambridge University Press},
  address = {Cambridge, UK and New York, NY, USA},
  year = {2022},
  pages = {3056},
  doi = {10.1017/9781009325844}
}

@misc{UNDP2024,
  author = {{United Nations Development Programme}},
  title = {A `Super Year' for Elections},
  year = {2024},
  howpublished = {\url{https://www.undp.org/super-year-elections}}
}

@article{TIMEElectionYear,
  author  = {Ewe, Koh},
  title   = {The Ultimate Election Year: All the Elections Around the World in 2024},
  journal = {TIME},
  year    = {2023},
  month   = {December},
  day     = {28},
  url     = {https://time.com/6550920/world-elections-2024/}
}

@article{schafer2026social,
  title={Social media and the changing landscape of climate change communication},
  author={Sch{\"a}fer, Mike S and Metag, Julia},
  journal={Nature Climate Change},
  volume={16},
  number={8},
  pages={897--906},
  year={2026},
  publisher={Nature Publishing Group UK London}
}

@article{misuraca2026energy,
  title={Energy transition as the core of climate discourse: Mapping social representations in digital public debate},
  author={Misuraca, Michelangelo and Spano, Maria and D’Aniello, Luca},
  journal={Energy Research \& Social Science},
  volume={140},
  pages={104945},
  year={2026},
  publisher={Elsevier}
}

@article{salloum2025anatomy,
  title={Anatomy of elite and mass polarization in social networks},
  author={Salloum, Ali and Chen, Ted Hsuan Yun and Kivel{\"a}, Mikko},
  journal={Network Science},
  volume={13},
  pages={e18},
  year={2025},
  publisher={Cambridge University Press}
}

@article{falkenberg2022growing,
  title={Growing polarization around climate change on social media},
  author={Falkenberg, Max and Galeazzi, Alessandro and Torricelli, Maddalena and Di Marco, Niccol{\`o} and Larosa, Francesca and Sas, Madalina and Mekacher, Amin and Pearce, Warren and Zollo, Fabiana and Quattrociocchi, Walter and others},
  journal={Nature Climate Change},
  volume={12},
  number={12},
  pages={1114--1121},
  year={2022},
  publisher={Nature Publishing Group UK London}
}

@article{falkenberg2026toward,
  title={Toward an integrated and globally equitable understanding of political polarization},
  author={Falkenberg, Max and Cinelli, Matteo and Galeazzi, Alessandro and Bail, Christopher A and Benito, Rosa M and Bruns, Axel and Gruzd, Anatoliy and Lazer, David and Lee, Jae Kook and McCoy, Jennifer and others},
  journal={Nature communications},
  volume={17},
  number={1},
  pages={8338},
  year={2026},
  publisher={Nature Publishing Group UK London}
}

@article{chen2021polarization,
  title={Polarization of climate politics results from partisan sorting: Evidence from Finnish Twittersphere},
  author={Chen, Ted Hsuan Yun and Salloum, Ali and Gronow, Antti and Yl{\"a}-Anttila, Tuomas and Kivel{\"a}, Mikko},
  journal={Global Environmental Change},
  volume={71},
  pages={102348},
  year={2021},
  publisher={Elsevier}
}

@article{loru2025ideology,
  title={Ideology and polarization set the agenda on social media},
  author={Loru, Edoardo and Galeazzi, Alessandro and Bonetti, Anita and Sangiorgio, Emanuele and Di Marco, Niccol{\`o} and Cinelli, Matteo and Falkenberg, Max and Baronchelli, Andrea and Quattrociocchi, Walter},
  journal={Scientific Reports},
  volume={15},
  number={1},
  pages={35816},
  year={2025},
  publisher={Nature Publishing Group UK London}
}

@article{scalco2026modelling,
  title={Modelling the energy transition debate in Italy through information supply and demand},
  author={Scalco, Irene and Colafrancesco, Giulia and Cinelli, Matteo},
  journal={npj Climate Action},
  year={2026},
  publisher={Nature Publishing Group UK London}
}

@article{torricelli2023does,
  title={How does extreme weather impact the climate change discourse? Insights from the Twitter discussion on hurricanes},
  author={Torricelli, Maddalena and Falkenberg, Max and Galeazzi, Alessandro and Zollo, Fabiana and Quattrociocchi, Walter and Baronchelli, Andrea},
  journal={Plos Climate},
  volume={2},
  number={11},
  pages={e0000277},
  year={2023},
  publisher={Public Library of Science San Francisco, CA USA}
}

@article{meyer2023between,
  title={Between calls for action and narratives of denial: Climate change attention structures on Twitter},
  author={Meyer, Hendrik and Peach, Amelia Katelin and Guenther, Lars and Kedar, Hadas Emma and Br{\"u}ggemann, Michael},
  journal={Media and Communication},
  volume={11},
  number={1},
  pages={278--292},
  year={2023},
  publisher={PRT}
}

@article{lochner2024climate,
  title={Climate summits and protests have a strong impact on climate change media coverage in Germany},
  author={Lochner, Jakob H and Stechemesser, Annika and Wenz, Leonie},
  journal={Communications Earth \& Environment},
  volume={5},
  number={1},
  pages={279},
  year={2024},
  publisher={Nature Publishing Group UK London}
}

@article{storani2025relative,
  title={Relative engagement with sources of climate misinformation is growing across social media platforms},
  author={Storani, Saverio and Falkenberg, Max and Quattrociocchi, Walter and Cinelli, Matteo},
  journal={Scientific Reports},
  volume={15},
  number={1},
  pages={18629},
  year={2025},
  publisher={Nature Publishing Group UK London}
}

@article{scalco2026detect,
  title={How to detect information voids using longitudinal data from social media and web searches},
  author={Scalco, Irene and Gesualdo, Francesco and Cerqueti, Roy and Cinelli, Matteo},
  journal={Scientific Reports},
  year={2026},
  publisher={Nature Publishing Group UK London}
}

@article{treen2020online,
  title={Online misinformation about climate change},
  author={Treen, Kathie M d'I and Williams, Hywel TP and O'Neill, Saffron J},
  journal={Wiley Interdisciplinary Reviews: Climate Change},
  volume={11},
  number={5},
  pages={e665},
  year={2020},
  publisher={Wiley Online Library}
}

@article{van2017inoculating,
  title={Inoculating the public against misinformation about climate change},
  author={Van der Linden, Sander and Leiserowitz, Anthony and Rosenthal, Seth and Maibach, Edward},
  journal={Global challenges},
  volume={1},
  number={2},
  pages={1600008},
  year={2017},
  publisher={Wiley Online Library}
}

@inproceedings{lan2024stance,
  title={Stance detection with collaborative role-infused llm-based agents},
  author={Lan, Xiaochong and Gao, Chen and Jin, Depeng and Li, Yong},
  booktitle={Proceedings of the international AAAI conference on web and social media},
  volume={18},
  pages={891--903},
  year={2024}
}

@article{dubey2025talking,
  title={Talking environment on TikTok: Messages, social actors, and engagement},
  author={Dubey, Shreya and Meijers, Marijn HC and Smit, Edith G and Smit, Eline S},
  journal={Environmental Communication},
  volume={19},
  number={4},
  pages={723--748},
  year={2025},
  publisher={Taylor \& Francis}
}

@article{hautea2021showing,
  title={Showing they care (or don’t): Affective publics and ambivalent climate activism on TikTok},
  author={Hautea, Samantha and Parks, Perry and Takahashi, Bruno and Zeng, Jing},
  journal={Social media+ society},
  volume={7},
  number={2},
  pages={20563051211012344},
  year={2021},
  publisher={SAGE Publications Sage UK: London, England}
}

@article{van2021elite,
  title={Elite influence on public attitudes about climate policy},
  author={Van Boven, Leaf and Sherman, David K},
  journal={Current Opinion in Behavioral Sciences},
  volume={42},
  pages={83--88},
  year={2021},
  publisher={Elsevier}
}

@article{cohen2003party,
  title={Party over policy: The dominating impact of group influence on political beliefs.},
  author={Cohen, Geoffrey L},
  journal={Journal of personality and social psychology},
  volume={85},
  number={5},
  pages={808},
  year={2003},
  publisher={American Psychological Association}
}

@article{rinscheid2021shapes,
  title={What shapes public support for climate change mitigation policies? The role of descriptive social norms and elite cues},
  author={Rinscheid, Adrian and Pianta, Silvia and Weber, Elke U},
  journal={Behavioural Public Policy},
  volume={5},
  number={4},
  pages={503--527},
  year={2021},
  publisher={Cambridge University Press}
}

@article{cinelli2020selective,
  title={Selective exposure shapes the Facebook news diet},
  author={Cinelli, Matteo and Brugnoli, Emanuele and Schmidt, Ana Lucia and Zollo, Fabiana and Quattrociocchi, Walter and Scala, Antonio},
  journal={PloS one},
  volume={15},
  number={3},
  pages={e0229129},
  year={2020},
  publisher={Public Library of Science San Francisco, CA USA}
}

@article{cinelli2021echo,
  title={The echo chamber effect on social media},
  author={Cinelli, Matteo and De Francisci Morales, Gianmarco and Galeazzi, Alessandro and Quattrociocchi, Walter and Starnini, Michele},
  journal={Proceedings of the national academy of sciences},
  volume={118},
  number={9},
  pages={e2023301118},
  year={2021},
  publisher={National Academy of Sciences}
}

@article{gilardi2022social,
  title={Social media and political agenda setting},
  author={Gilardi, Fabrizio and Gessler, Theresa and Kubli, Ma{\"e}l and M{\"u}ller, Stefan},
  journal={Political communication},
  volume={39},
  number={1},
  pages={39--60},
  year={2022},
  publisher={Taylor \& Francis}
}

@article{weng2012competition,
  title={Competition among memes in a world with limited attention},
  author={Weng, Lilian and Flammini, Alessandro and Vespignani, Alessandro and Menczer, Fillipo},
  journal={Scientific reports},
  volume={2},
  number={1},
  pages={335},
  year={2012},
  publisher={Nature Publishing Group UK London}
}

@online{europarl_topics,
  author = {{European Parliament}},
  title = {Topics},
  publisher = {European Parliament},
  url = {https://www.europarl.europa.eu/topics/en}
}

@article{gerlach2018network,
  title={A network approach to topic models},
  author={Gerlach, Martin and Peixoto, Tiago P and Altmann, Eduardo G},
  journal={Science advances},
  volume={4},
  number={7},
  pages={eaaq1360},
  year={2018},
  publisher={American Association for the Advancement of Science}
}

@article{lorenz2019accelerating,
  title={Accelerating dynamics of collective attention},
  author={Lorenz-Spreen, Philipp and M{\o}nsted, Bjarke M{\o}rch and H{\"o}vel, Philipp and Lehmann, Sune},
  journal={Nature communications},
  volume={10},
  number={1},
  pages={1759},
  year={2019},
  publisher={Nature Publishing Group UK London}
}

@article{sangiorgio2025evaluating,
  title={Evaluating the effect of viral posts on social media engagement},
  author={Sangiorgio, Emanuele and Di Marco, Niccol{\`o} and Etta, Gabriele and Cinelli, Matteo and Cerqueti, Roy and Quattrociocchi, Walter},
  journal={Scientific Reports},
  volume={15},
  number={1},
  pages={639},
  year={2025},
  publisher={Nature Publishing Group UK London}
}

@misc{Meta_Report,
  author = {{Meta}},
  title = {Regulation (EU) 2022/2065 Digital Services Act Transparency Report for Facebook},
  year = {2024},
  url = {https://transparency.meta.com/reports/regulatory-transparency-reports/}
}

@article{hartland2026role,
  title={The role of key European issues in the 2024 election campaign},
  author={Hartland, Alex and Braun, Daniela and Carteny, Giuseppe and Navarrete, Rosa M and Reinl, Ann-Kathrin},
  journal={West european politics},
  volume={49},
  number={1},
  pages={265--292},
  year={2026},
  publisher={Taylor \& Francis}
}

@article{kluver2016responds,
  title={Who responds? Voters, parties and issue attention},
  author={Kl{\"u}ver, Heike and Spoon, Jae-Jae},
  journal={British Journal of Political Science},
  volume={46},
  number={3},
  pages={633--654},
  year={2016},
  publisher={Cambridge University Press}
}

@article{kluver2016setting,
  title={Setting the agenda or responding to voters? Political parties, voters and issue attention},
  author={Kl{\"u}ver, Heike and Sagarzazu, I{\~n}aki},
  journal={West European Politics},
  volume={39},
  number={2},
  pages={380--398},
  year={2016},
  publisher={Taylor \& Francis}
}

@article{ennser2022does,
  title={Does social media enhance party responsiveness? How user engagement shapes parties’ issue attention on Facebook},
  author={Ennser-Jedenastik, Laurenz and Gahn, Christina and Bodlos, Anita and Haselmayer, Martin},
  journal={Party Politics},
  volume={28},
  number={3},
  pages={468--481},
  year={2022},
  publisher={SAGE Publications, Sage UK: London, England}
}

@article{farstad2018explains,
  title={What explains variation in parties’ climate change salience?},
  author={Farstad, Fay M},
  journal={Party politics},
  volume={24},
  number={6},
  pages={698--707},
  year={2018},
  publisher={SAGE Publications, Sage UK: London, England}
}

@article{roxburgh2019characterising,
  title={Characterising climate change discourse on social media during extreme weather events},
  author={Roxburgh, Nicholas and Guan, Dabo and Shin, Kong Joo and Rand, William and Managi, Shunsuke and Lovelace, Robin and Meng, Jing},
  journal={Global environmental change},
  volume={54},
  pages={50--60},
  year={2019},
  publisher={Elsevier}
}

@article{hart2025climate,
  title={Climate change advocacy and engagement on social media},
  author={Hart, P Sol and Feldman, Lauren and Choi, Soobin and Chinn, Sedona and Hiaeshutter-Rice, Dan},
  journal={Science Communication},
  volume={47},
  number={3},
  pages={385--412},
  year={2025},
  publisher={Sage Publications, Sage CA: Los Angeles, CA}
}

@article{barbera2019leads,
  title={Who leads? Who follows? Measuring issue attention and agenda setting by legislators and the mass public using social media data},
  author={Barber{\'a}, Pablo and Casas, Andreu and Nagler, Jonathan and Egan, Patrick J and Bonneau, Richard and Jost, John T and Tucker, Joshua A},
  journal={American Political Science Review},
  volume={113},
  number={4},
  pages={883--901},
  year={2019},
  publisher={Cambridge University Press}
}

@article{team2025gemma,
  title={Gemma 3 technical report},
  author={Team, Gemma and Kamath, Aishwarya and Ferret, Johan and Pathak, Shreya and Vieillard, Nino and Merhej, Ramona and Perrin, Sarah and Matejovicova, Tatiana and Ram{\'e}, Alexandre and Rivi{\`e}re, Morgane and others},
  journal={arXiv preprint arXiv:2503.19786},
  year={2025}
}

@techreport{lafortune2026europe,
  author = {Lafortune, Guillaume and Fuller, Grayson},
  title = {Europe Sustainable Development Report 2026: {SDG} Pathways to 2030 and Mid-Century},
  institution = {SDSN and Dublin University Press},
  address = {Paris and Dublin},
  year = {2026},
  doi = {10.82163/224},
  note = {DOI: \url{https://doi.org/10.82163/224}}
}

@article{shim2020measuring,
  title={Measuring the mass-elite preference congruence: Findings from a meta-analysis and introduction to the symposium},
  author={Shim, Jaemin and Gherghina, Sergiu},
  journal={European Political Science},
  volume={19},
  number={4},
  pages={509--527},
  year={2020},
  publisher={Cambridge University Press}
}

@article{yoda2026time,
  title={Time dilation: the temporal structure of elite--voter incongruence on redistribution and immigration},
  author={Yoda, Hironao and Atsumi, Serika},
  journal={West European Politics},
  pages={1--29},
  year={2026},
  publisher={Taylor \& Francis}
}

@article{rovny20252024,
  title={The 2024 Chapel Hill Expert Survey on political party positioning in Europe: Twenty-five years of party positional data},
  author={Rovny, Jan and Polk, Jonathan and Bakker, Ryan and Hooghe, Liesbet and Jolly, Seth and Marks, Gary and Steenbergen, Marco and Vachudova, Milada Anna},
  journal={Electoral Studies},
  volume={97},
  pages={102981},
  year={2025},
  publisher={Elsevier}
}

\clearpage
\includepdf[pages=-]{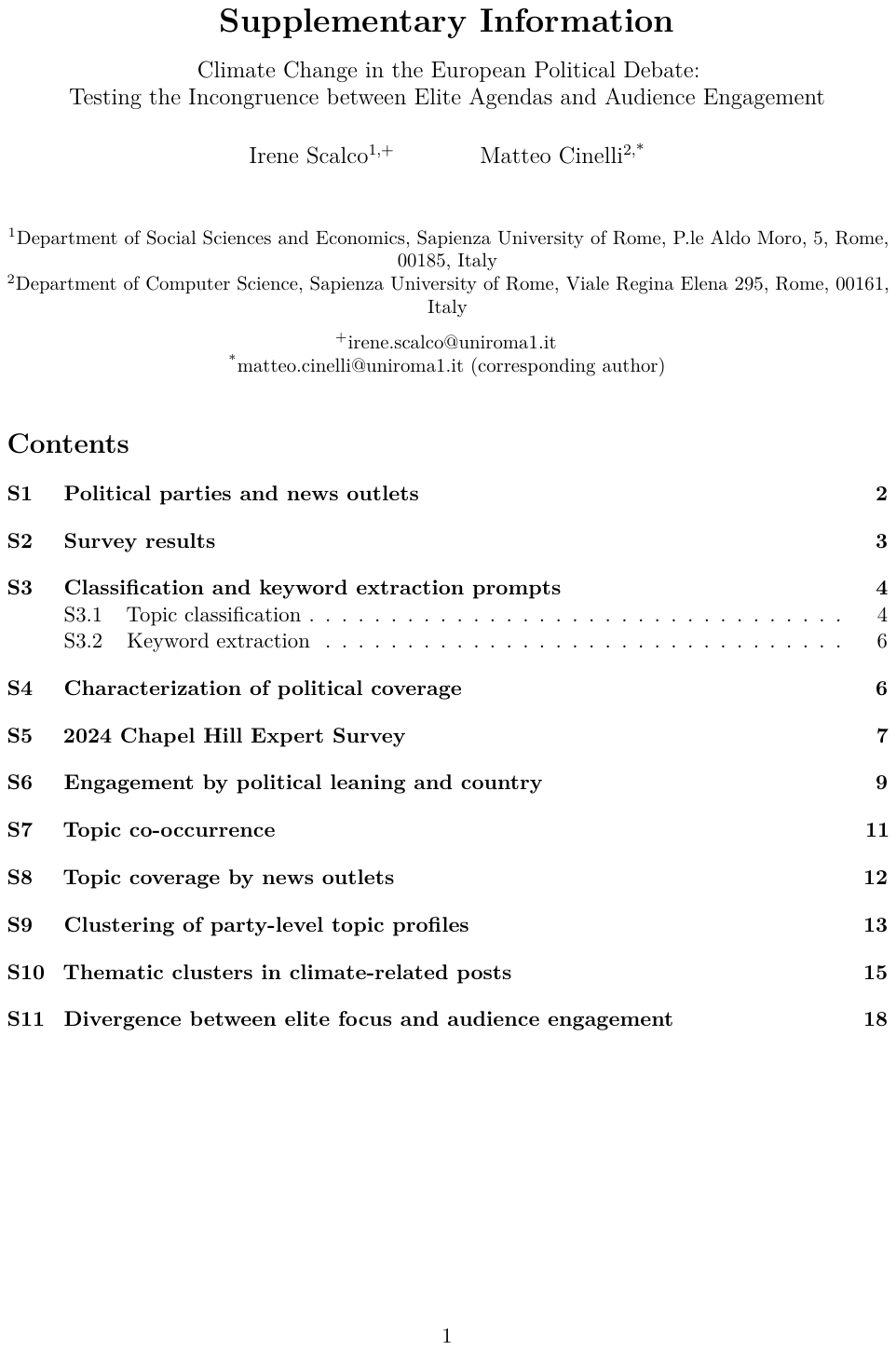} 
\end{document}